\documentclass[conference]{IEEEtran}
\IEEEoverridecommandlockouts
\usepackage{amsmath,amssymb,amsfonts}
\usepackage{booktabs}
\usepackage{algorithmic}
\usepackage{graphicx}
\usepackage{textcomp}
\usepackage{siunitx}
\usepackage{graphicx}
\usepackage[hidelinks]{hyperref}
\usepackage{url} 
\usepackage{tikz}
\usepackage{float}
\usepackage{xcolor}
\usepackage[font=footnotesize,labelsep=period]{caption}

\def\BibTeX{{\rm B\kern-.05em{\sc i\kern-.025em b}\kern-.08em
    T\kern-.1667em\lower.7ex\hbox{E}\kern-.125emX}}

\newcommand{\subtitle}[1]{\noindent\textbf{{#1}}}

\usepackage[xindy, toc, numberedsection, acronym]{glossaries}
\glsdisablehyper

\newacronym{2d}{2D}{two-dimensional}
\newacronym{3d}{3D}{three-dimensional}
\newacronym{3gpp}{3GPP}{3rd generation partnership project}
\newacronym{4g}{4G}{fourth-generation}
\newacronym{5g}{5G}{fifth-generation}
\newacronym{6dof}{6DoF}{six degrees of freedom}
\newacronym{6g}{6G}{sixth-generation}
\newacronym{abp}{ABP}{authentication by personalisation}
\newacronym{ac}{AC}{alternating current}
\newacronym{aclr}{ACLR}{adjacent channel leakage ratio}
\newacronym{adc}{ADC}{analog-to-digital converter}
\newacronym{adr}{ADR}{adaptive data rate}
\newacronym{afe}{AFE}{analog front end}
\newacronym{ag}{AG}{array gain}
\newacronym{ai}{AI}{artificial intelligence}
\newacronym{am}{AM}{amplitude modulation}
\newacronym{amam}{AM/AM}{amplitude modulation to amplitude modulation}
\newacronym{amc}{AMC}{automatic modulation classification}
\newacronym{amp}{AMP}{approximate message passing}
\newacronym{ampm}{AM/PM}{amplitude modulation to phase modulation}
\newacronym{aoa}{AOA}{angle-of-arrival}
\newacronym{aod}{AOD}{angle-of-departure}
\newacronym{ap}{AP}{access point}
\newacronym{api}{API}{application program interface}
\newacronym{apu}{APU}{access point unit}
\newacronym{ar}{AR}{augmented reality}
\newacronym{arima}{ARIMA}{Auto-Regressive Integrated Moving Average}
\newacronym{arp}{ARP}{Antenna Reference Point}
\newacronym{asic}{ASIC}{application specific integrated circuit}
\newacronym{ask}{ASK}{amplitude-shift keying}
\newacronym{auv}{AUV}{autonomous underwater vehicle}
\newacronym{awgn}{AWGN}{additive white Gaussian noise}
\newacronym{baw}{BAW}{bulk acoustic wave}
\newacronym{bb}{BB}{base-band}
\newacronym{bc}{BC}{backscatter communication}
\newacronym{bd}{BD}{backscatter device}
\newacronym{be}{BE}{Belgium}
\newacronym{ber}{BER}{bit error rate}
\newacronym{bf}{BF}{beamforming}
\newacronym{bldc}{BLDC}{brushless DC}
\newacronym{ble}{BLE}{Bluetooth Low Energy}
\newacronym{bom}{BOM}{bill of materials}
\newacronym{bpsk}{BPSK}{binary phase-shift keying}
\newacronym{bs}{BS}{base station}
\newacronym{bu}{BU}{booster unit}
\newacronym{bw}{BW}{bandwidth}
\newacronym{ca}{CA}{carrier emitter}
\newacronym{cad}{CAD}{channel activity detection}
\newacronym{cars}{CARS}{calibration reference signal}
\newacronym{cbm}{CBM}{condition based maintenance}
\newacronym{cc}{CC}{constant current}
\newacronym{ccdf}{CCDF}{complementary cumulative distribution function}
\newacronym{ccnn}{CCNN}{circular convolutional neural network}
\newacronym{ccs}{CCS}{correlative channel sounder}
\newacronym{cdf}{CDF}{cumulative distribution function}
\newacronym{cdrx}{CDRX}{connected mode DRX}
\newacronym{ce}{CE}{coverage enhancement}
\newacronym{ced}{CED}{cumulative energy density}
\newacronym{cf}{CF}{cell-free}
\newacronym{cfmmimo}{CF-mMIMO}{cell-free massive MIMO}
\newacronym{cfo}{CFO}{carrier frequency offset}
\newacronym{cir}{CIR}{channel impulse response}
\newacronym{cla}{CLA}{closed-loop approach}
\newacronym{cmos}{CMOS}{complementary metal oxide semiconductor}
\newacronym{cost}{COST}{commercial off-the-shelf}
\newacronym{cots}{COTS}{commercial off-the-shelf}
\newacronym{cp}{CP}{cyclic prefix}
\newacronym{cpt}{CPT}{capacitive power transfer}
\newacronym{cpu}{CPU}{central-processing unit}
\newacronym{cqi}{CQI}{channel quality indicator}
\newacronym{cr}{CR}{coding rate}
\newacronym{crc}{CRC}{cyclic redundancy check}
\newacronym{crlb}{CRLB}{Cram\'er-Rao lower bound}
\newacronym{crs}{CRS}{cell reference signal}
\newacronym{cs}{CS}{compressed sensing}
\newacronym{cse}{CSE}{channel state estimation}
\newacronym{csi}{CSI}{channel state information}
\newacronym{csp}{CSP}{contact service point}
\newacronym{css}{CSS}{chirp spread spectrum}
\newacronym{cu}{CU}{central unit}
\newacronym{cv}{CV}{constant voltage}
\newacronym{cw}{CW}{continuous wave}
\newacronym{d2d}{D2D}{device-to-device}
\newacronym{dab}{DAB}{Digital Audio Broadcasting}
\newacronym{dac}{DAC}{digital-to-analog converter}
\newacronym{daq}{DAQ}{data acquisition system}
\newacronym{das}{DAS}{distributed antenna systems}
\newacronym{dc}{DC}{direct current}
\newacronym{dcc}{DCC}{dynamic cooperation clustering}
\newacronym{dcu}{DCU}{Data Capturing Unit}
\newacronym{de}{DE}{drain efficiency}
\newacronym{dft}{DFT}{discrete Fourier transform}
\newacronym{dl}{DL}{downlink}
\newacronym{dlc}{DLC}{Distributed Laser Charging}
\newacronym{dli}{DLI}{direct link interference}
\newacronym{dm}{DM}{diffuse multipath}
\newacronym{dma}{DMA}{Direct Memory Access}
\newacronym{dmc}{DMC}{diffuse multipath component}
\newacronym{dmimo}{D-MIMO}{distributed MIMO}
\newacronym{doa}{DOA}{direction-of-arrival}
\newacronym{dpd}{DPD}{digital pre-distortion}
\newacronym{dpdk}{DPDK}{Data Plane Development Kit}
\newacronym{dram}{DRAM}{dynamic random-access memory}
\newacronym{drcs}{$\Delta$RCS}{differential-radar cross section}
\newacronym{drx}{DRX}{Discontinuous Reception Mode}
\newacronym{dsl}{DSL}{digital subscriber line}
\newacronym{dsp}{DSP}{digital signal processing}
\newacronym{dss}{DSS}{Dataset Storage Standard}
\newacronym{duc}{DUC}{digital up-converter}
\newacronym{dvb}{DVB}{Digital Video Broadcasting}
\newacronym{e2e}{E2E}{end-to-end}
\newacronym{easa}{EASA}{European Union Aviation Safety Agency}
\newacronym{ec}{EC}{European Commission}
\newacronym{ecdf}{eCDF}{empirical cumulative distribution function}
\newacronym{ecsp}{ECSP}{edge computing service point}
\newacronym{edlc}{EDLC}{electrostatic double-layer capacitors}
\newacronym{edrx}{eDRX}{Extended Discontinuous Reception Mode}
\newacronym{ee}{EE}{energy efficiency}
\newacronym{ef}{EF}{Environmental Footprint}
\newacronym{egprs}{EGPRS}{Enhanced Data Rates for GSM Evolution}
\newacronym{eh}{EH}{energy harvesting}
\newacronym{eirp}{EIRP}{equivalent isotropically radiated power}
\newacronym{em}{EM}{electromagnetic}
\newacronym{embb}{eMBB}{enhanced Mobile Broadband}
\newacronym{en}{EN}{energy neutral}
\newacronym{end}{END}{energy neutral device}
\newacronym{enob}{ENOB}{effective number of bits}
\newacronym{eol}{EoL}{end-of-life}
\newacronym{epu}{EPU}{edge processing unit}
\newacronym{er}{ER}{Energy Receiver}
\newacronym{erp}{ERP}{equivalent radiated power}
\newacronym[plural=ESCs,firstplural=electronic speed controllers (ESCs)]{esc}{ESC}{electronic speed control}
\newacronym{esd}{ESD}{electrostatic discharge}
\newacronym{esl}{ESL}{electronic shelf label}
\newacronym{esr}{ESR}{equivalent series resistance}
\newacronym{et}{ET}{Energy Transmitter}
\newacronym{etsi}{ETSI}{European Telecommunications Standards Institute}
\newacronym{evd}{EVD}{eigenvalue decomposition}
\newacronym{evm}{EVM}{Error Vector Magnitude}
\newacronym{ewlb}{eWLB}{Embedded Wafer Level Ball Grid Array}
\newacronym{fa}{FA}{federation anchor}
\newacronym{fair}{FAIR}{Findability, Accessibility, Interoperability, and Reuse of digital assets}
\newacronym{fd}{FD}{front-haul distance}
\newacronym{fde}{FDE}{frequency domain equalizer}
\newacronym{fdm}{FDM}{frequency-division multiplexing}
\newacronym{fdma}{FDMA}{frequency division multiple access}
\newacronym{fem}{FEM}{finite element analysis}
\newacronym{fembb}{feMBB}{further enhanced mobile broadband}
\newacronym{fft}{FFT}{fast Fourier transform}
\newacronym{fh}{FH}{fronthaul}
\newacronym{fifo}{FIFO}{First In, First Out}
\newacronym{fim}{FIM}{Fisher information matrix}
\newacronym{fits}{FITS}{Flexible Image Transport System}
\newacronym{fom}{FoM}{figure of merit}
\newacronym{fpga}{FPGA}{field-programmable gate array}
\newacronym{fsk}{FSK}{frequency shift keying}
\newacronym{fsr}{FSR}{Full Scale Range}
\newacronym{fss}{FSS}{frequency selective surface}
\newacronym{gb}{GB}{grant-based}
\newacronym{gf}{GF}{grant-free}
\newacronym{gmsk}{GMSK}{Gaussian minimum-shift keying}
\newacronym{gnb}{gNB}{Next Generation Node B}
\newacronym{gni}{GNI}{gross national income}
\newacronym{gnn}{GNN}{graph neural network}
\newacronym{gnss}{GNSS}{global navigation satellite system}
\newacronym{gpclk}{GPCLK}{general purpose clock}
\newacronym{gpl}{GPL}{GNU General Public License}
\newacronym{gprs}{GPRS}{General Packet Radio Services}
\newacronym{gps}{GPS}{Global Positioning System}
\newacronym{gpu}{GPU}{graphical processing unit}
\newacronym{grc}{GRC}{GNU Radio Companion}
\newacronym{gscm}{GSCM}{geometry‐based stochastic model}
\newacronym{gsm}{GSM}{Global System for Mobile Communications}  %
\newacronym{gwp}{GWP}{Global Warming Potential}
\newacronym{harq}{HARQ}{hybrid automatic repeat request}
\newacronym{hat}{HAT}{hardware attached on top}
\newacronym{hcs}{HCS}{human-centric services}
\newacronym{hdf5}{HDF5}{Hierarchical Data Format version 5}
\newacronym{hfss}{HFSS}{High Frequency Simulator Software}
\newacronym{hmd}{HMD}{head-mounted display}
\newacronym{hpbm}{HPBM}{half power beam width}
\newacronym{HyMPRo}{HyMPRo}{Hybrid Multi-Path Routing algorithm}
\newacronym{i.i.d.}{i.i.d.}{independent and identically distributed}
\newacronym{ib}{IB}{in-band}
\newacronym{ibo}{IBO}{input back-off}
\newacronym{ic}{IC}{integrated circuit}
\newacronym{ici}{ICI}{intercarrier interference}
\newacronym{id}{ID}{information decoding}
\newacronym{idft}{IDFT}{inverse discrete Fourier transform}
\newacronym{if}{IF}{intermediate-frequency}
\newacronym{iid}{i.i.d.}{independently and identically distributed}
\newacronym{iiot}{IIoT}{Industrial Internet of Things}
\newacronym{iis}{IIS}{integrated information system}
\newacronym{im}{IM}{intermodulation}
\newacronym{imu}{IMU}{inertial measurement unit}
\newacronym{io}{IO}{input/output}
\newacronym{iot}{IoT}{Internet of Things}
\newacronym{ipt}{IPT}{inductive power transfer}
\newacronym{ipy}{IPY}{Interventions per Year}
\newacronym{iq}{IQ}{in-phase and quadrature}
\newacronym{iqi}{IQI}{IQ imbalance}
\newacronym{ir}{IR}{infrared}
\newacronym{isi}{ISI}{intersymbol interference}
\newacronym{ism}{ISM}{industrial, scientific and medical}
\newacronym{isp}{ISP}{internet service provider}
\newacronym{jesd}{JESD}{Joint Electron Devices Engineering Council}
\newacronym{jfet}{JFET}{junction field effect transistor}
\newacronym{kpi}{KPI}{key performance indicator}
\newacronym{kvi}{KVI}{key value indicator}
\newacronym{larva}{LARVA}{LARge Virtual Array}
\newacronym{lca}{LCA}{life-cycle assessment}
\newacronym{lcia}{LCIA}{life cycle impact assessment}
\newacronym{lco}{LCO}{lithium cobalt oxide}
\newacronym{ldo}{LDO}{Low-dropout voltage regulator}
\newacronym{ldpc}{LDPC}{low-density parity-check}
\newacronym{less}{LESS}{Low Energy Scheduler Solution}
\newacronym{lfp}{LFP}{lithium iron phosphate}
\newacronym{lic}{LIC}{lithium-ion capacitor}
\newacronym{lidar}{LiDAR}{light detection and ranging}
\newacronym{liion}{Li-ion}{lithium-ion}
\newacronym{lipo}{LiPo}{lithium polymer}
\newacronym{lis}{LIS}{large intelligent surface}
\newacronym{llh}{LLH}{log-likelihood}
\newacronym{lmmse}{LMMSE}{least minimum mean square error}
\newacronym{lmo}{LMO}{lithium ion manganese oxide}
\newacronym{lna}{LNA}{low-noise amplifier}
\newacronym{lo}{LO}{local oscillator}
\newacronym{lora}{LoRa}{long range}
\newacronym{lorawan}{LoRaWAN}{long-range wide-area network}
\newacronym{los}{LoS}{line-of-sight}
\newacronym{lp}{LP}{linear programming}
\newacronym{lpt}{LPT}{laser power transfer}
\newacronym{lpwa}{LPWA}{Low Power Wide Area}
\newacronym{lpwan}{LPWAN}{low-power wide-area network}
\newacronym{lqi}{LQI}{link quality indicator}
\newacronym{lrelu}{LReLU}{leaky rectified linear unit}
\newacronym{lrt}{LRT}{likelihood-ratio test}
\newacronym{ls}{LS}{least squares}
\newacronym{lsa}{LSA}{large synthetic array}
\newacronym{lsf}{LSF}{large-scale fading}
\newacronym{lsfc}{LSFC}{large-scale fading component}
\newacronym{lstm}{LSTM}{Long Short-Term Memory}
\newacronym{ltc}{LTC}{lithium thionyl chloride}
\newacronym{lte}{LTE}{Long Term Evolution}
\newacronym{lti}{LTI}{linear time-invariant}
\newacronym{lto}{LTO}{lithium titanate}
\newacronym{lusta}{LUSTA 5G }{Logistique mUltimodale Sécuritaire Téléopérée \& Autonome 5G}
\newacronym{m2m}{M2M}{machine to machine}
\newacronym{mac}{MAC}{Medium Access Control}
\newacronym{mate}{MATE}{millimeter-wave MIMO testbed}
\newacronym{mc}{MC}{Monte Carlo}
\newacronym{mcl}{MCL}{Maximum Coupling Loss}
\newacronym{mcs}{MCS}{modulation and coding scheme}
\newacronym{mcu}{MCU}{Microcontroller Unit}
\newacronym{mec}{MEC}{multi-access edge computing}
\newacronym{mems}{MEMS}{micro-electromechanical systems}
\newacronym{mf}{MF}{matched filter}
\newacronym{mimo}{MIMO}{multiple-input multiple-output}
\newacronym{miso}{MISO}{multiple-input single-output}
\newacronym{ml}{ML}{machine learning}
\newacronym{mlp}{MLP}{multilayer perceptron}
\newacronym{mmimo}{mMIMO}{massive MIMO}
\newacronym{mmse}{MMSE}{minimum mean square error}
\newacronym{mmtc}{mMTC}{massive machine-typed communication}
\newacronym{mmwave}{mmWave}{millimeter wave}
\newacronym{mn}{MN}{matching network}
\newacronym{mosfet}{MOSFET}{metal-oxide semiconductor field effect transistor}
\newacronym{mpc}{MPC}{multipath component}
\newacronym{mppt}{MPPT}{maximum power point tracking}
\newacronym{mr}{MR}{maximum ratio}
\newacronym{mrc}{MRC}{maximum ratio combining}
\newacronym{mrc_em}{MRC}{maximum ratio combining}
\newacronym{mrc_EM}{MRC}{Magnetic Resonance Coupling}
\newacronym{mrt}{MRT}{maximum ratio transmission}
\newacronym{mse}{MSE}{mean square error}
\newacronym{mtc}{MTC}{Machine-Type Communication}
\newacronym{multi-rat}{Multi-RAT}{multiple radio access technology}
\newacronym{music}{MUSIC}{MUltiple SIgnal Classification}
\newacronym{navauwall}{NAVAUWALL}{AUtomated NAVigation in WALLonia}
\newacronym{nb}{NB}{narrowband}
\newacronym{nbiot}{NB-IoT}{narrowband IoT}
\newacronym{nca}{NCA}{nickel cobalt aluminum}
\newacronym{netcdf}{NetCDF}{Network Common Data Form}
\newacronym{nfc}{NFC}{near-field communication}
\newacronym{nfv}{NFV}{network function virtualization}
\newacronym{ngmn}{NGMN}{Next Generation Mobile Networks }
\newacronym{ni}{NI}{National Instruments}
\newacronym{nicd}{NiCd}{nikkel cadmium}
\newacronym{nimh}{NiMH}{nikkel metal hydride}
\newacronym{nlos}{NLoS}{non-line-of-sight}
\newacronym{nmc}{NMC}{nickel manganese cobalt}
\newacronym{nmos}{nMOS}{n-channel metal-oxide semiconductor}
\newacronym{nn}{NN}{neural network}
\newacronym{nnls}{NNLS}{non-negative least squares}
\newacronym{noma}{NOMA}{non-orthogonal multiple access}
\newacronym{np}{NP}{Neyman-Pearson}
\newacronym{npbch}{NPBCH}{Narrowband Physical Broadcast Channel}
\newacronym{npss}{NPSS}{Narrow Band Primay Synchronization Signal}
\newacronym{nr}{NR}{New Radio}
\newacronym{nrs}{NRS}{Narrow Band Reference Signal}
\newacronym{nsss}{NSSS}{Narrowband Secondary Synchronization Signal}
\newacronym{ntp}{NTP}{network time protocol}
\newacronym{oai}{OAI}{OpenAirInterface} 
\newacronym{obw}{OBW}{occupied bandwidth}
\newacronym{ofdm}{OFDM}{orthogonal frequency-division multiplexing}
\newacronym{ofdma}{OFDMA}{orthogonal frequency-division multiple access}
\newacronym{ofdmim}{OFDM-IM}{OFDM with index modulation}
\newacronym{oob}{OOB}{out-of-band}
\newacronym{ook}{OOK}{on-off keying}
\newacronym{oran}{O-RAN}{open radio-access network}
\newacronym{os}{OS}{operating system}
\newacronym{ota}{OTA}{over-the-air}
\newacronym{otaa}{OTAA}{over-the-air authentication}
\newacronym{p1}{P1}{Phase 1}
\newacronym{p2}{P2}{Phase 2}
\newacronym{p2p}{P2P}{point-to-point}
\newacronym{pa}{PA}{power amplifier}
\newacronym{pae}{PAE}{power-added efficiency}
\newacronym{pana}{PanA}{Panel A}
\newacronym{panb}{PanB}{Panel B}
\newacronym{papr}{PAPR}{peak-to-average power ratio}
\newacronym{pc}{PC}{pilot count}
\newacronym{pcb}{PCB}{printed circuit board}
\newacronym{pcg}{PCG}{power consumption gain}
\newacronym{pcie}{PCIe}{Peripheral Component Interconnect Express}
\newacronym{pcsi}{PCSI}{perfect channel state information}
\newacronym{pd}{PD}{powered device}
\newacronym{pdcch}{PDCCH}{physical downlink control channel}
\newacronym{pdf}{PDF}{probability density function}
\newacronym{pdp}{PDP}{power delay profile}
\newacronym{pdsch}{PDSCH}{physical downlink shared channel}
\newacronym{peb}{PEB}{positioning error bound}
\newacronym{per}{PER}{packet error rate}
\newacronym{pet}{PET}{polyethylene terephthalate}
\newacronym{pg}{PG}{path gain}
\newacronym{pgd}{PGD}{proximal gradient descent}
\newacronym{phy}{PHY}{physical}
\newacronym{pl}{PL}{path loss}
\newacronym{pll}{PLL}{phase-locked loop}
\newacronym{pmf}{PMF}{polymer microwave fiber}
\newacronym{poe}{PoE}{power-over-Ethernet}
\newacronym{pps}{1PPS}{pulse per second}
\newacronym{prbs}{PRBs}{Physical Resource Blocks}
\newacronym{ps}{PS}{Processing System}
\newacronym{psd}{PSD}{power spectral density}
\newacronym{pse}{PSE}{power sourcing equipment}
\newacronym{psk}{PSK}{phase shift keying}
\newacronym{psm}{PSM}{power saving mode}
\newacronym{pss}{PSS}{primary synchronisation signal}
\newacronym{ptp}{PTP}{precision-time protocol}
\newacronym{ptrs}{PTRS}{Phase-Tracking Reference Signals}
\newacronym{ptw}{PTW}{paging time window}
\newacronym{pv}{PV}{photovoltaic}
\newacronym{pw}{PW}{planar wavefront}
\newacronym{pwm}{PWM}{pulse width modulation}
\newacronym{qam}{QAM}{quadrature amplitude modulation}
\newacronym{qos}{QoS}{quality-of-service}
\newacronym{qpsk}{QPSK}{quadrature phase-shift keying}
\newacronym{quadriga}{QuaDRiGa}{QUAsi Deterministic RadIo channel GenerAtor}
\newacronym{ra}{RA}{Random Access}
\newacronym{ram}{RAM}{random-access memory}
\newacronym{ran}{RAN}{radio access network}
\newacronym{rar}{RAR}{Random Access Response}
\newacronym{rat}{RAT}{radio access technology}
\newacronym{rb}{Rb}{Rubidium}
\newacronym{rbs}{RBS}{radio base station}
\newacronym{rbw}{RBW}{resolution bandwidth}
\newacronym{rcs}{RCS}{radar cross section}
\newacronym{re}{RE}{radio element}
\newacronym{relu}{ReLU}{rectified linear unit}
\newacronym{rf}{RF}{radio frequency}
\newacronym{rfeh}{RFEH}{radio frequency energy harvesting}
\newacronym{rfic}{RFIC}{radio-frequency integrated circuit}
\newacronym{rfid}{RFID}{radio frequency identification}
\newacronym{rfpt}{RFPT}{radio frequency power transfer}
\newacronym{rfsoc}{RFSoC}{Radio Frequency System-on-Chip}
\newacronym{rir}{RIR}{room impulse response}
\newacronym{ris}{RIS}{reflective intelligent surface}
\newacronym{rllmtc}{RLLMTC}{reliable low latency machine type communication}
\newacronym{rls}{RLS}{recursive least squares}
\newacronym{rms}{RMS}{root-mean-square}
\newacronym{rmse}{RMSE}{root-mean-square error}
\newacronym{rmt}{RMT}{random matrix theory}
\newacronym{rof}{RoF}{radio-over-fiber}
\newacronym{ros}{ROS}{robot operating system}
\newacronym{rpi}{RPi}{Raspberry Pi}
\newacronym{rrc}{RRC}{Radio Resource Connection}
\newacronym{rreq}{RREQ}{route request packet}
\newacronym{rsrp}{RSRP}{Reference Signals Received Power}
\newacronym{rsrq}{RSRQ}{Reference Signal Received Quality}
\newacronym{rss}{RSS}{received signal strength}
\newacronym{rssi}{RSSI}{received signal strength indicator}
\newacronym{rtc}{RTC}{real time clock}
\newacronym{rtk}{RTK}{real time kinematics}
\newacronym{rts}{RTS}{ray tracing simulator}
\newacronym{ru}{RU}{Radio Unit}
\newacronym{rv}{RV}{random variable}
\newacronym{rw}{RW}{RadioWeaves}
\newacronym{rx}{RX}{receiver}
\newacronym{rzf}{RZF}{regularized zero forcing}
\newacronym{s-parameter}{S-parameter}{scattering parameter}
\newacronym{sa}{SA}{synchronization anchor}
\newacronym{sa5}{SA}{Stand Alone}
\newacronym{sar}{SAR}{specific absorption rate}
\newacronym{sbl}{SBL}{sparse Bayesian learning}
\newacronym{scfdma}{SCFDMA}{single-carrier frequency division multiple access}
\newacronym{sdg}{SDG}{Sustainable Development Goal}
\newacronym{sdm}{SDM}{sigma-delta modulator}
\newacronym{sdn}{SDN}{software-defined network}
\newacronym{sdof}{SDoF}{sigma-delta over fiber}
\newacronym{sdr}{SDR}{software-defined radio}
\newacronym{se}{SE}{spectral efficiency}
\newacronym{sei}{SEI}{specific emitter identification}
\newacronym{ser}{SER}{symbol-error rate}
\newacronym{sf}{SF}{spreading factor}
\newacronym{sfn}{SFN}{single frequency network}
\newacronym{sfp}{SFP}{small form-factor pluggable}
\newacronym{SigMF}{SigMF}{Signal Metadata Format}
\newacronym{simo}{SIMO}{single-input multiple-output}
\newacronym{sinr}{SINR}{signal-to-interference-plus-noise ratio}
\newacronym{siso}{SISO}{single-input single-output}
\newacronym{slam}{SLAM}{simultaneous localization and mapping}
\newacronym{slc}{SLC}{spatial leakage suppression}
\newacronym{slerp}{SLERP}{spherical linear interpolation}
\newacronym{sma}{SMA}{SubMiniature version A}
\newacronym{smc}{SMC}{specular multipath component}
\newacronym{smps}{SMPS}{switched mode power supply}
\newacronym{sndr}{SNDR}{signal-to-noise-and-distortion ratio}
\newacronym{snidr}{SNIDR}{signal-to-noise-and-interference-and-distortion ratio}
\newacronym{snr}{SNR}{signal-to-noise ratio}
\newacronym{soc}{SoC}{state of charge}
\newacronym{soc2}{SoC}{system-on-chip}
\newacronym{sram}{SRAM}{static random-access memory}
\newacronym{srd}{SRD}{short-range device}
\newacronym{srs}{SRS}{Sounding Reference Signal}
\newacronym{ssb}{SSB}{synchronisation signal block}
\newacronym{ssd}{SSD}{solid state drive}
\newacronym{ssq}{SSQ}{simulator sickness questionnaire}
\newacronym{steam}{STEAM}{science, technology, engineering, the arts, and mathematics}
\newacronym{svd}{SVD}{singular value decomposition}
\newacronym{sw}{SW}{spherical wavefront}
\newacronym{swipt}{SWIPT}{simultaneous wireless information and power transfer}
\newacronym{synce}{SyncE}{Synchronous Ethernet}
\newacronym{tau}{TAU}{tracking area update}
\newacronym{tcer}{TCER}{transported to consumed energy ratio}
\newacronym{tcp}{TCP}{Transmission Control Protocol}
\newacronym{tdd}{TDD}{time division duplexing}
\newacronym{tdoa}{TDOA}{time-difference-of-arrival}
\newacronym{thz}{THz}{Terahertz}
\newacronym{to}{TO}{timing offset}
\newacronym{toa}{TOA}{time-of-arrival}
\newacronym{tof}{ToF}{time-of-flight}
\newacronym{tosm}{TOSM}{through-open-short-match}
\newacronym{trl}{TRL}{technology readyness level}
\newacronym{trp}{TRP}{Transmission Reception Point}
\newacronym{tsn}{TSN}{time-sensitive networking}
\newacronym{ttm}{TTM}{time to market}
\newacronym{ttn}{TTN}{The Things Network}
\newacronym{tx}{TX}{transmitter}
\newacronym{uav}{UAV}{unmanned aerial vehicle}
\newacronym{uc}{UC}{use case}
\newacronym{ucie}{UCIe}{Universal Chiplet Interconnect Express}
\newacronym{udp}{UDP}{User Datagram Protocol}
\newacronym{ue}{UE}{user equipment}
\newacronym{ugv}{UGV}{unmanned ground vehicle}
\newacronym{uhd}{UHD}{USRP hardware driver}
\newacronym{uhf}{UHF}{ultra-high frequency}
\newacronym{uhs}{UHS}{Ultra High Speed}
\newacronym{ul}{UL}{uplink}
\newacronym{ula}{ULA}{uniform linear array}
\newacronym{uMIMO}{$\mu$-MIMO}{ultra-massive Multiple-Input Multiple-Output}
\newacronym{ummtc}{umMTC}{ultra massive machine type communication}
\newacronym{un}{UN}{United Nations}
\newacronym{upa}{UPA}{uniform planar array}
\newacronym{ura}{URA}{uniform rectangular array}
\newacronym{urllc}{URLLC}{ultra-reliable low-latency communications}
\newacronym{usrp}{USRP}{universal software radio peripheral}
\newacronym{uv}{UV}{unmanned vehicle}
\newacronym{uwb}{UWB}{ultrawideband}
\newacronym{v2v}{V2V}{vehicle-to-vehicle}
\newacronym{vep}{VEP}{virtual edge platform}
\newacronym{vlc}{VLC}{visible light communication}
\newacronym{vlp}{VLP}{visible light positioning}
\newacronym{vna}{VNA}{vector network analyzer}
\newacronym{votable}{VOTable}{Virtual Observatory Table}
\newacronym{vr}{VR}{virtual reality}
\newacronym{wb}{WB}{wideband}
\newacronym{wban}{WBAN}{wireless body area network}
\newacronym{wimax}{WiMAX}{Worldwide Interoperability for Microwave Access}
\newacronym{wlan}{WLAN}{wireless LAN}
\newacronym{wom}{WOM}{wake-on-motion}
\newacronym{wpt}{WPT}{wireless power transfer}
\newacronym{wr}{WR}{White Rabbit}
\newacronym{wrsn}{WRSN}{wireless rechargeable sensor network}
\newacronym{wsn}{WSN}{wireless sensor network}
\newacronym{xets}{XETS}{cross exponentially tapered slot}
\newacronym{xr}{XR}{extended reality}
\newacronym{z3ro}{Z3RO}{zero third-order distortion}
\newacronym{zf}{ZF}{zero-forcing}
\newacronym{zmcscg}{ZMCSCG}{zero mean circularly symmetric complex Gaussian}

\providecommand{\gls}[1]{#1}

\providecommand{\acrshort}[1]{#1}

\ifCLASSINFOpdf
\else
\fi

\begin{document}

\title{Low-Power Impact Detection and Localization \\on Forklifts Using Wireless IMU Sensors\\}
\author{
    \IEEEauthorblockN{Lyssa Ramaut, Chesney Buyle, Jona Cappelle, and Liesbet Van der Perre}
    \IEEEauthorblockA{\textit{KU Leuven, ESAT-WaveCore},
    \textit{Ghent Technology Campus}, Ghent, Belgium \\ 
    name.surname@kuleuven.be}
    \vspace{-0.8cm}
}


\maketitle
\pagestyle{empty}
\begin{abstract}

Forklifts are essential for transporting goods in industrial environments. These machines face wear and tear during field operations, along with rough terrain, tight spaces and complex handling scenarios. This increases the likelihood of unintended impacts, such as collisions with goods, infrastructure, or other machinery. In addition, deliberate misuse has been stated, compromising safety and equipment integrity. This paper presents a low-cost and low-power impact detection system based on multiple wireless sensor nodes measuring 3D accelerations. These were deployed in a measurement campaign covering real-world operational scenarios. An algorithm was developed, based on this collected data, to differentiate high-impact events from normal usage and to localize detected collisions on the forklift. The solution successfully detects and localizes impacts, while maintaining low power consumption, enabling reliable forklift monitoring with multi-year sensor autonomy.  

\end{abstract}

\begin{IEEEkeywords}
impact detection algorithm, collision localization, accelerometers, wireless sensor network, remote monitoring
\vspace{-0.4cm}

\end{IEEEkeywords}


\section{Introduction}

Forklifts are often leased to companies for years and then returned for maintenance. Consequently, rental companies have limited insight into how the equipment is used, making damage or improper handling detection difficult. This creates a need for monitoring systems that can detect, classify, and report abnormal usage like impacts or misuse. Such systems enhance safety and accountability, support proactive maintenance, and help reduce long-term operational costs.



Many reported solutions rely on a single accelerometer with a G\mbox{-}force threshold to detect impact. For example, ~\cite{lehoczky-2022} classifies events over \SI{10}{G} as impacts, while ~\cite{yee-2018} differentiates hard braking (\SI{1}{G}-\SI{2}{G}) from crashes ($>$\SI{20}{G}). Other low-complexity systems are based on tilt angles~\cite{faiz-2015, chaudhary-2020}. 
Advanced systems combine low and high-G accelerometers to cover a wide range of impacts, along with gyroscopes for angular changes and a GPS for velocity and position data~\cite{phuong2018degree}.
However, these mainly detect crashes rather than localize impacts. To address this, some methods add sensors like ultrasonic distance sensors and cameras to assess both the severity and position of impacts~\cite{rana-2020}. 
Importantly, most threshold-based systems in literature target normal or high-speed vehicles,  where relying solely on deceleration or tilt patterns is effective. Forklifts, however, experience low-speed impacts requiring different detection approaches. In addition, recent work also explores \acrshort{ai}-driven methods for industrial vehicles. Commercial solutions such as SIERA.AI S2~\cite{sieraai-2024} and iWAREHOUSE Impact Management~\cite{IwareHouse} distinguish collisions from normal driving, but rely on large datasets, constant cloud connectivity and high power consumption.
    
In summary, the state of the art lacks a robust, low-power impact detection and localization system that can be easily retrofitted to industrial vehicles. It should detect and localize impacts, with minimal cabling, supporting easy deployment and flexible sensor placement across various vehicle types. 
    
This paper presents a novel system comprising low-complexity, in-house developed wireless accelerometer nodes and a tailored algorithmic framework for impact detection and collision localization. A dataset of real-world forklift operations, covering both routine and impact scenarios, is collected and used to validate the system's performance. The overall design emphasizes plug-and-play installation, ultra-low power operation, and robustness to sensor misalignment through an effective calibration procedure. 
Section~II outlines the system setup, data collection procedure and calibration process. In Section~III, the collision detection algorithm is described, followed by a discussion of the results in Section~IV. The power consumption of the accelerometer nodes is discussed in Section~V. Final conclusions are provided in Section~VI.
\begin{figure}[t]
\centering
\includegraphics[width=\columnwidth]{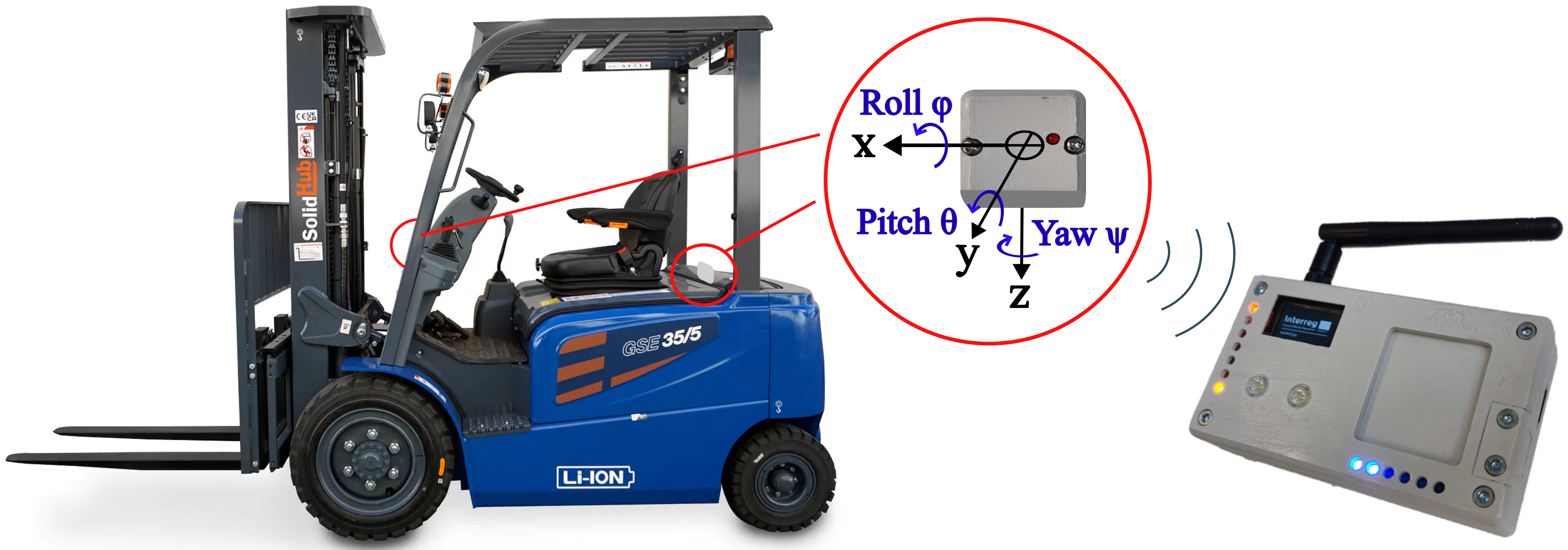} \vspace{-0.5cm}
\caption{Forklift setup with two IMUs and central data capturing unit (DCU)}
\vspace{-0.1cm}
\label{fig:fig_acc_setup}
\end{figure}
\section{System, data collection and calibration}
\subsection{System overview}
The system consists of two low-power in-house developed wireless sensor nodes, each built around a 9-DoF ICM-20948 \gls{imu} \cite{unknown-author-2024}. These communicate with a central \gls{dcu} via \gls{ble} and synchronization accuracy below \SI{200}{\micro\second}~\cite{cappelle2024lowpower}. The design, calibration functions and other specifications of this platform are elaborated further in~\cite{cappelle-2020}. We set up a dedicated measurement campaign in which one sensor node was mounted at the front and the other at the back of the forklift, as shown in Fig.~\ref{fig:fig_acc_setup}. The accelerometers sample data at \si{100}{Hz} with a \gls{fsr} of \SI{8}{G}. The data sent to the \gls{dcu} from the two sensor nodes is read out and analyzed on a computer. 
Our goal was to investigate whether clear and reliable event-related patterns could be found using this low-complexity sensing system.




\subsection{Dataset: collection and calibration}

Data were collected at TVH in Waregem (Belgium), a company that sells, rents, and maintains off-road machinery. A ‘Still RX20-20’ forklift equipped with the in-house monitoring system was used. The dataset includes everyday tasks like loading trucks and driving on uneven surfaces, as well as deliberate misuse with controlled collisions at the left and right back corners. Table \ref{tab:results_all} summarizes this dataset.
To enable easy installation and compensate for any misalignments during sensor mounting, a calibration procedure is required to transform the accelerometer data from a tilted sensor frame (T) to a leveled reference frame (L). To explain this procedure, we first derive the relationship between these frames to estimate the misalignment angles \textbf{roll $\phi$}, \textbf{pitch $\theta$} and \textbf{yaw $\psi$}, whose orientations are shown in Fig.~\ref{fig:fig_acc_setup}. During start-up, this calibration automatically corrects for these former two using the 3-2-1 Euler rotation matrix, given by:
\vspace{-0.1cm}
\begin{flalign*}
&R^{T}_{L} = \ 
\begin{bmatrix}
c_\theta c_\psi & s_\psi c_\theta & -s_\theta \\
s_\phi s_\theta c_\psi - c_\phi s_\psi & s_\phi s_\theta s_\psi + c_\phi c_\psi & c_\theta s_\phi \\
s_\theta c_\phi c_\psi + s_\phi s_\psi & s_\theta s_\psi c_\phi - c_\psi s_\phi & c_\theta c_\phi
\end{bmatrix} \\
&\text{with } c_x = \cos x, \quad s_x = \sin x \quad\text{and} \quad x \in \{\phi, \theta, \psi\}.
\end{flalign*}
When the forklift is not moving, the accelerations in the leveled plane are known $P^{L} = [0 \quad 0 \quad g]^T$. To express these in the sensor's (possibly misaligned) tilted frame, the rotation matrix $R^{T}_{L}$ is applied: $P^{T}=R^{T}_{L} \cdot P^{L}$.
Consequently, the measured acceleration components for $P^{T}$ are:
{\setlength{\abovedisplayskip}{4pt}
\setlength{\belowdisplayskip}{4pt}
\begin{equation*}
    a_x = -g \sin\theta, \ 
    a_y = g \cos\theta \sin\phi \text{ and } 
    a_z = g \cos\theta \cos\phi\text{.}
\end{equation*}}
Since $a_{x}$, $a_{y}$ and $a_{z}$ are the (known) measured sensor values, the \textbf{pitch} $\theta$ and \textbf{roll} $\phi$ angles can be computed as:
{\setlength{\abovedisplayskip}{6pt}
\setlength{\belowdisplayskip}{4pt}
\begin{equation*}
\theta = \arcsin\left(\frac{a_x}{-g}\right) \ 
\quad\text{and}\quad \phi = \arctan\left(\frac{a_y}{a_z}\right)\text{.} 
\end{equation*}}
The 3-2-1 Euler rotation matrix $R^T_L(\phi, \theta)$ is used (with $\psi$ = $0^\circ$) to transform the measured accelerations $P^{T} = [a_x \quad a_y \quad a_z ]^T$ into the leveled frame accelerations $P^L$: 
{\setlength{\abovedisplayskip}{0.05cm}
\setlength{\belowdisplayskip}{0.1cm}
\begin{equation*}
P^L = (R^T_L)^{-1} \cdot P^T
\end{equation*}}
The \textbf{yaw} $\psi$ angle on the other hand, cannot be determined from static accelerometer data alone. However, tests show that small yaw errors have minimal impact on the algorithm, hence clear sensor installation guidelines should be sufficient. For larger yaw misalignments, a rough estimate of $\psi$ can be calculated when the forklift is in motion: 
{\setlength{\abovedisplayskip}{4pt}
\setlength{\belowdisplayskip}{4pt}
\begin{equation*}
\psi = \arctan\left(\frac{a_y}{a_x}\right)
\end{equation*}}
Tests, however, revealed that the relatively small acceleration components were highly susceptible to driving noise, leading to significant variability in the estimated yaw angle.
Consequently this method should primarily serve to detect large misalignment (e.g.,  $>$90$^{\circ}$ or  $>$180$^{\circ}$ rotation) and notify the installer or driver of incorrect sensor installation.\\
\renewcommand{\arraystretch}{0.9} 
{\small
\vspace{-1.1cm}
\begin{table*}[h!]
\captionsetup{singlelinecheck=on}
\centering \caption{Validation results for all tested events and detections.\\ BT = below threshold, AT! = above threshold (severe), LB = left back, RB = right back.}
\vspace{-0.2cm}
\centering
\resizebox{\textwidth}{!}{%
\begin{tabular}{@{}l|l|l@{}}

\toprule
\textbf{Datafile.log} & \textbf{Event} & \textbf{Detection}      \\ 
\midrule
\textbf{Collision events (LB/RB)} & & \\
collision\_rb.log     & 5x RB (very soft, soft, hard, harder, hard)  & 1x short vibration BT, 3x RB collision, 1x short vibration BT \\
collision\_lb.log     & 5x LB (hard, hard, hard, soft, hard) & 1x short vibration BT, 3x LB collision, 2x short vibration BT   \\ 
\midrule
\textbf{Short \& long vibration events} & & \\
driving\_bumpy\_road.log        & Normal driving behavior             & 10x short/long vibrations BT, 1x long vibration AT!     \\
driving\_bumpy\_road(2).log    & Intentionally taking bumps fast  & 2x braking, 5x short/long vibration AT!, 1x short vibration BT \\
loading\_truck.log     & Normally loading a truck & 10x short vibrations BT \\
\midrule
\textbf{Harsh braking events} & & \\
normal\_braking.log        & Soft braking             & Nothing                            \\
hard\_braking.log     & 2x harsh braking         & 2x braking                        \\
hard\_braking(2).log     & 3x very harsh braking         & 3x braking, 1x short vibration BT \\
\midrule
\textbf{Undetectable events} & & \\
picking\_up\_load.log       & Brutally picking up a load             & Nothing                            \\
forks\_up\_down.log     & Desliberate fork-ground contact         & 1x short vibration BT \\
\end{tabular}%
}
\vspace{-0.5cm}
\label{tab:results_all}
\end{table*}
}
 \vspace{0.7cm}\section{Impact detection algorithm}
\begin{figure}[h]
\vspace{-0.6cm}
\centering
\includegraphics[width=\columnwidth]{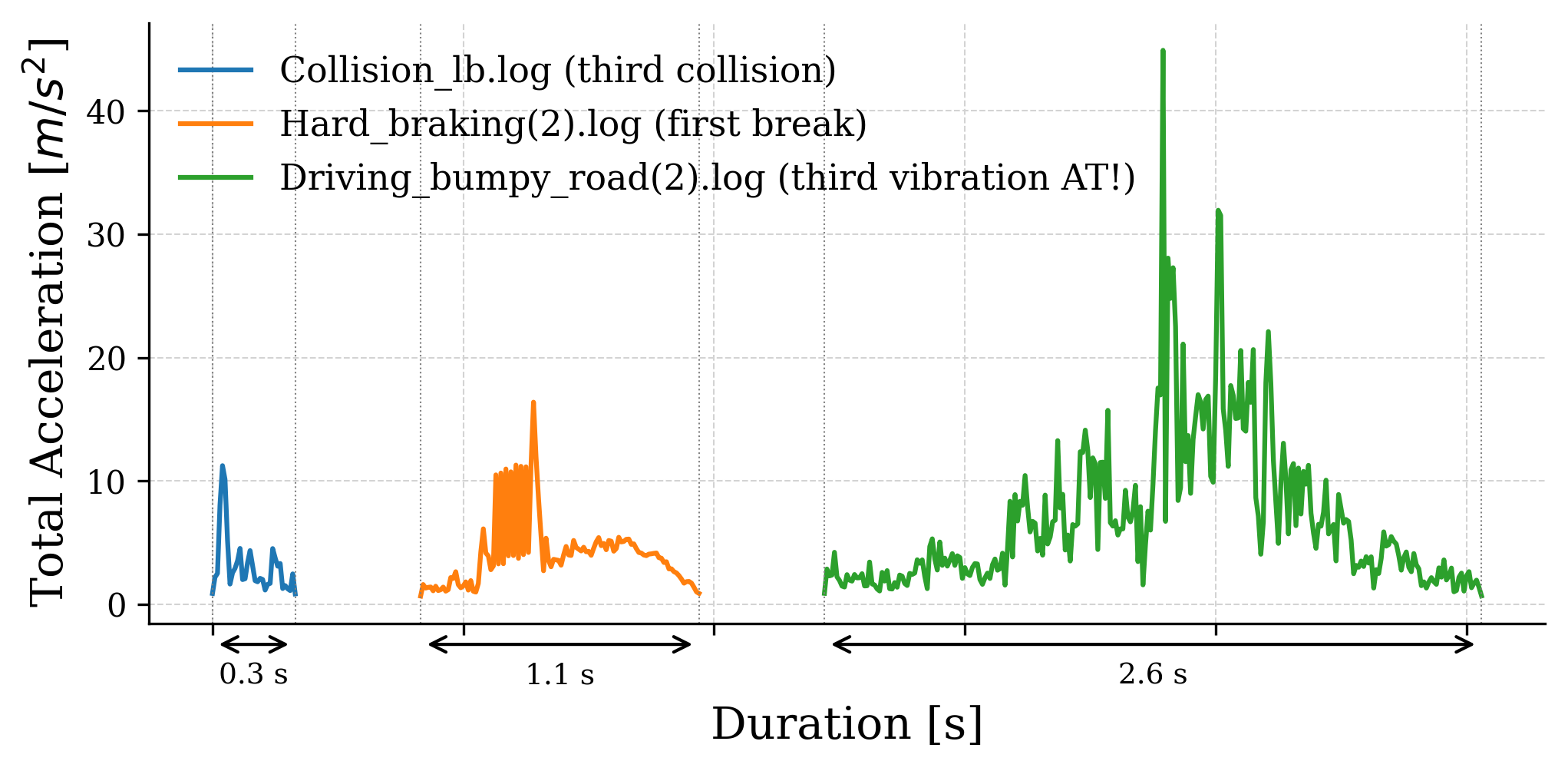} \vspace{-0.65cm}
\caption{Total acceleration of all sensors (mean) for different events}
\label{fig_Alg_board_1}
\vspace{0.15cm}
\end{figure}

 \vspace{-0.1cm}
A multi-stage algorithm was used to detect different impact types, including harsh braking, collisions, and improper driving. 
First, high-impact events are detected using thresholds and categorized by features such as duration and signal shape. 
These were carefully selected through dataset analysis and comparative evaluation. Example signals visualized in Fig.~\ref{fig_Alg_board_1} help clarify some of these thresholds.  The different algorithm steps are detailed below. \\
\\
\subtitle{Event detection and segment extraction:} The total acceleration \( a_{\text{total}} = \sqrt{a_x^2 + a_y^2 + a_z^2} \) is calculated for the front and back sensor, and then averaged across both to obtain \( a_{\text{total, mean}}\). Events that cause peaks above a threshold of \SI{5}{m/s^{2}} trigger segment extraction, which are extended until the signal drops below \SI{1}{m/s^{2}}. To avoid fragmentation, adjacent segments with \SI{0.5}{s} in between are merged and only segments longer than \SI{5}{ms} are retained for further analysis.

\begin{figure}[h]
\vspace{-0.3cm}
\centering
\includegraphics[width=0.9\columnwidth]{
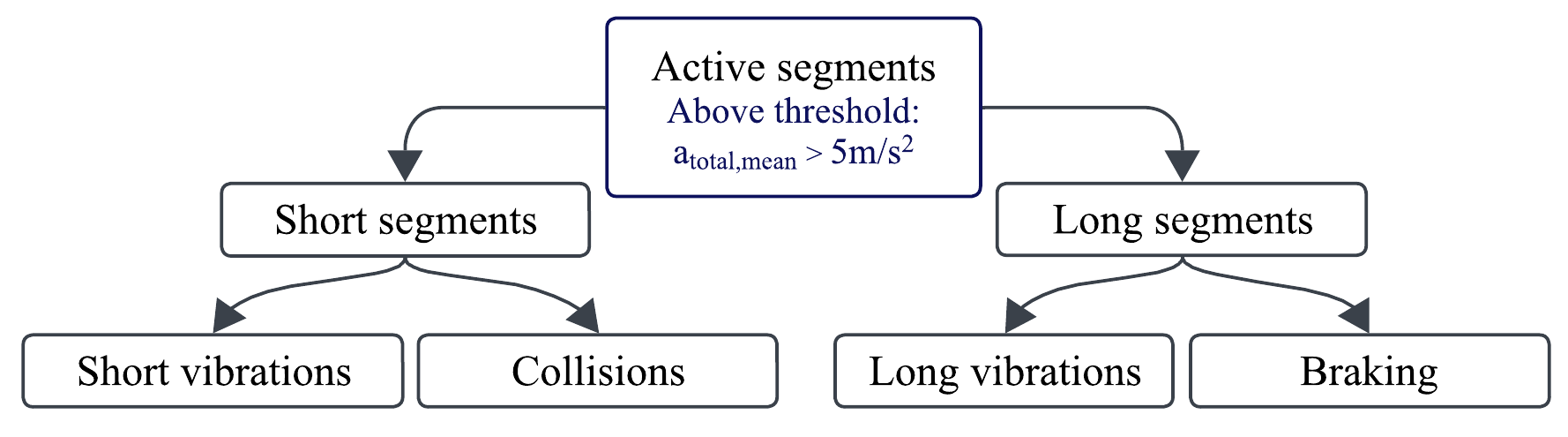} 
\vspace{-0.2cm}
\caption{Algorithm overview: event categorization}
\label{fig_Alg_board_0}
\vspace{0.15cm}
\end{figure}

\subtitle{Segment categorization:}
Extracted segments are categorized by duration and signal features, as shown in Fig.~\ref{fig_Alg_board_0}. 
Segments with durations between \SI{5}{ms} and \SI{750}{ms} are considered short, those longer than \SI{1.25}{s} long. Intermediate durations are classified by the ratio of net to total area under the \( a_x \) curve to identify breaks early, enabling accurate classification later based on signal features. More precisely, braking causes strong deceleration, keeping the signal mostly on one side of the baseline and yielding a high ratio. Vibrations and collisions cause fluctuations, resulting in a lower ratio. If the ratio exceeds $75\%$, the segment is marked long, if not short. 
 

Short segments are categorized by comparing the total area under the \( a_y \) and \( a_z \) curves of the sensor with the highest absolute \( a_y \) value. Focusing on this sensor prevents averaging effects that could blur collision accelerations, as these signals are weaker for sensors further from the impact and may be masked then by vibrations. If \( a_y \) is greater than \( a_z \), the segment is labeled as a collision; otherwise as a vibration. This matches physics: collisions typically cause stronger acceleration parallel to ground (\( a_y \)), while vibrations from uneven surfaces mainly affect the vertical axis (\( a_z\)).
Long segments, on the other hand, are classified as braking or vibrations based on the number of baseline crossings, as braking has less crossings than vibrations.  
    
\subtitle{Collision localization: }
Additionally, the collision location (front or back) is determined by the sensor with the highest peak in \( a_y \). More precisely, this \( a_y \) axis is chosen for its clear impact peaks compared to \( a_x\), which often also contains normal longitudinal accelerations (e.g., braking). The sign of the net \( a_y \) then indicates the side of impact: positive indicates a right-side impact (left movement), negative a left-side impact. This maps collisions to four zones: left front, right front, left back or right back.\\
\subtitle{Harsh braking: }
Braking events with the total deceleration exceeding \SI{5}{m/s^2} are labeled as harsh. To distinguish from acceleration, the net area under the \( a_y \) curve is used. Negative area confirms braking and a positive area indicates acceleration. This avoids false positives from sudden starts.\\
\subtitle{Short and long vibrations: }
Vibrations are classified as normal and abnormal using a threshold of \SI{22}{m/s^2} on \( a_{\text{total, mean}} \). Segments above this threshold are labeled AT! (above threshold), others as BT (below threshold). This filters out routine operations like truck loading, while identifying abnormal driving like driving fast over bumps. Events like abrupt load pickup or fork operation show weak, inconsistent signals, making their detection unreliable with only two sensors.
\vspace{-0.15cm}
\section{Results and Discussion}


The proposed system,  deploying two sensors and a multi-stage detection algorithm, successfully detects and classifies high-impact forklift events, as shown by Table \ref{tab:results_all}. Specifically, medium and hard collisions are correctly localized within one of the four predefined zones. In addition, the algorithm effectively distinguishes normal driving behavior (e.g., driving over uneven surfaces, loading a truck or normal braking) from misuse scenarios like taking bumps at high speed. However, some low-impact events and impacts with inconsistent signal patterns (e.g., very soft collisions or fork impacts) remain undetected or are classified as below-threshold vibrations.

Despite its robustness, the current algorithm relies on fixed thresholds for acceleration magnitude, segment duration, and baseline crossings. Those were chosen based on experimental data. However further investigation is required to determine their generalizability across different forklift models and environments. For example, other surfaces like dirt or gravel may alter vibration patterns and affect the performance. Therefore, further testing with a larger, diverse dataset is needed to assess the algorithm’s consistency and robustness on unseen data.

\section{Power consumption}
The wireless nodes use a small battery and integrate an nRF52832 \gls{soc2} with \gls{ble} connectivity alongside their \gls{imu}. They continuously measure and transmit data to the \gls{dcu}. However, to achieve multi-year autonomy, a \gls{wom}-based approach is needed, triggering sensor measurements when an acceleration threshold is exceeded. Two modes were evaluated: \textit{active mode} with continuous sampling and data transmission and \textit{sleep mode} with \gls{wom} activation. Power consumption for each mode is:
\begin{itemize}
    \item \textbf{Sleep mode} (\gls{wom} enabled): \SI{82.4}{\micro\watt}
    \item \textbf{Active mode} (continuous sampling + BLE): \SI{27.2}{\milli\watt}
\end{itemize}

Assuming a \SI{15}{Wh} battery pack (e.g., 4 AA batteries), the estimated autonomy when using \gls{wom} strongly depends on the number of daily activations:
\begin{itemize}
    \item \textbf{Low activity} (720 triggers/day): 8.8 years autonomy
    \item \textbf{High activity} (5000 triggers/day): 2 years autonomy
\end{itemize}

While \gls{wom} can extend sensor autonomy, its practical application in collision detection is limited by hardware constraints. In particular, the \gls{imu} used in this study takes up to \SI{20}{ms} to wake up from sleep mode, missing critical acceleration peaks. Consequently, an IMU with faster \gls{wom} or continuously powered \gls{imu} should be used. 

\section{Conclusion}
This paper presents a low-power, wireless multi-sensor \gls{imu} system with a tailored, threshold-based algorithm for detecting and localizing impacts on forklifts. The system reliably identifies medium and hard collisions within one of four predefined zones, while also accurately classifying vibrations and braking events between normal and misuse scenarios. 
Future work will focus on improving generalization with adaptive thresholds to ensure robustness across multiple vehicles, and power-saving strategies.
\section*{Acknowledgment}
We would like to thank THV and GemOne (Waregem, Belgium) for identifying the industrial need and the collaboration on this project. Their support, particularly in providing access to their forklift, was invaluable in collecting sufficient data.
\bibliography{bib}
\end{document}